\documentclass[aps,pre,amssymb,amsmath]{revtex4}
\usepackage[dvips]{graphicx}
\usepackage{amssymb}
\usepackage{hhline}

\newcommand{\be}{\begin{equation}}
\newcommand{\ee}{\end{equation}}
\newcommand{\bea}{\begin{eqnarray}}
\newcommand{\eea}{\end{eqnarray}}
\newcommand{\p}{\partial}

\newcommand{\iv}{\mathbf{i}}
\newcommand{\vvect}{\mathbf{v}}
\newcommand{\mydiv}{\mathrm{div}\,}
\newcommand{\mygrad}{{\pmb\nabla}}

\newcommand{\rot}{\mathrm{rot}\,}
\newcommand{\w}{\varphi}
\newcommand{\ta}{\theta}
\begin{document}
\title{Influence of gravitational forces and fluid flows \\
on a shape of surfaces of a viscous fluid of capillary size}
\author{L.Yu.~Barash}
\affiliation{
Landau Institute for Theoretical Physics,
142432 Chernogolovka, Russia\\
e-mail: \tt barash@itp.ac.ru
}
\begin{abstract}
The Navier-Stokes equations and boundary conditions
for viscous fluids of capillary size are formulated in
curvilinear coordinates associated with a
geometry of the fluid-gas interface. As a result,
the fluid dynamics of drops and menisci can be described
taking into account an influence of gravitational forces and flows
on the surface shape. This gives a convenient
basis for respective numerical studies. Estimations of the effects
are presented for the case of an evaporating sessile drop.
\end{abstract}
\maketitle

{\it Introduction.}
A number of important physical features in studying fluid flows
in evaporating liquid drops and menisci of capillary size has been
found recently both theoretically and experimentally~\cite{Deegan97,
Deegan,HuLarsonEvap,Popov1,HuLarsonMarangoni,Girard,Ristenpart,Dhaval}.
In particular, it was demonstrated that the vortex convection takes place
in evaporating drops and menisci
under various conditions~\cite{HuLarsonMarangoni,Girard,Ristenpart,Dhaval}. 
The activity in the field is associated now
with important applications. The particular examples are 
the evaporative contact line deposition~\cite{Deegan97,Deegan,Govor,
HuLarsonCoffee,Popov1,Popov2}, studies of DNA stretching behavior
and DNA mapping methods~\cite{Jing,HuLarsonDNA,Hsieh},
developing methods for jet ink printing~\cite{Park,Jong,Lim},
self-assembly of nanocrystal superlattice 
monolayer~\cite{Lin1,Lin2,Bigioni3}.

For describing the processes theoretically one should
carry out, in general, a joint study of the fluid dynamics, the
thermal conduction and the vapor diffusion together with respective
boundary conditions, in particular at the fluid-gas interface.
Standard approximations used in the theoretical studies are
a spherical cap shape of the drop or menisci and
a neglection of the hydrodynamical pressure and velocities in
the generalized Laplace formula. Though such approximations
can be justified under certain conditions, there are a wide range of 
parameters of the problem when a more accurate theoretical description
of liquid surfaces of capillary size is needed.

A shape of a surface is, generally, controlled
by combined effects of a surface tension, gravitational forces,
a hydrodynamic pressure and a~velocity distribution near the surface.
For solving fluid dynamics problems in a vicinity of curved
surfaces of an arbitrary shape, an explicit approach
is developed in the present paper, making use
of ``natural'' curvilinear coordinates associated with
a surface geometry. Both fluid dynamics equations and
the respective boundary conditions are formulated in
these coordinates. The equations in such a form are
convenient for numerical simulations.
We also present analytical estimations for the effects
in question for the case of an evaporating sessile drop,
which follow from the obtained results.

{\it Equations and boundary conditions.} The Navier-Stokes equations take the form
\be
\frac{\p\vvect}{\p t}+(\vvect\nabla)\vvect+
\frac1\rho\,{\mathrm{grad}\,p}
=\nu\,\Delta{\vvect}.
\label{NavSt}
\ee
For simplicity, we assume below that a shape of the surface as well as
fluid flows are axially symmetric and $v_\ta=0$, where $r$, $\ta$,
$z$ are cylindrical coordinates. This property is valid for a wide class
of problems. Therefore, it is convenient to use cylindrical coordinates
in the bulk of an incompressible viscous liquid and to introduce vorticity
$\gamma=\p v_r/\p z-\p v_z/\p r$ and the stream function $\psi$, such that
$\p \psi/\p z=r v_r$, $\p\psi/\p r=-r v_z$
(as distinct from the two-dimensional case~\cite{LL6}).
Then $\rot\vvect=\gamma(r,z)\mathbf{i}_\ta$,
the continuity equation $\mydiv \mathbf{v}=0$ is naturally satisfied.
Equations for quantities $\gamma$, $\psi$ are given by
\begin{eqnarray}
\frac{\p}{\p t}\gamma(r,z)+(\vvect\nabla)\gamma(r,z)&=&
\nu \left(\Delta\gamma(r,z)-\frac{\gamma(r,z)}{r^2}\right),\\
\Delta \psi-\frac{2}{r}\frac{\p\psi}{\p r}&=&r\gamma.
\end{eqnarray}
In order to formulate equations close to the surface,
it is convenient to choose orthogonal curvilinear coordinates
$x^n(x,y,z)$, $x^\tau(x,y,z)$, $x^{\theta}(x,y,z)$
with local basis vectors normal and tangential to
the surface at every point. In order to write down in these
curvilinear coordinates the differential forms
which enter the hydrodynamic equations, one needs to find
explicit expressions for the metric tensor and Christoffel symbols
for the chosen class of coordinate systems. Consider both
the contravariant coordinates $x^n$, $x^\tau$, $x^{\theta}$ and
the respective physical curvilinear coordinates $n,\tau,\tau_\ta$.
Locally $dn$ is a length along the normal to the surface,
$d\tau$ is a surface arc length in the meridian plane,
and $d\tau_\ta$ is a surface arc length associated with the rotation
angle around the $z$ axis. For an axially symmetric surface
$d\tau_\ta=r(n,\tau)d\theta$. For a differential of radius-vector
we have
\begin{equation}
d\mathbf{r}=dx^n \mathbf{e}_n+dx^\tau \mathbf{e}_\tau+
dx^\ta \mathbf{e}_\ta=dn\,\mathbf{i}_n+d\tau\,\mathbf{i}_\tau+
rd\ta\,\mathbf{i}_\ta\, ,
\end{equation}
where $\mathbf{e}_\ell=\mathbf{i}\,{dx}/{dx^\ell}+
\mathbf{j}\,{dy}/{dx^\ell}+\mathbf{k}\,{dz}/{dx^\ell}$ are
contravariant base vectors.
Unlike contravariant base vectors
$\mathbf{e}_n,\mathbf{e}_\tau,\mathbf{e}_\ta$,
Cartesian base vectors
$\mathbf{i},\mathbf{j},\mathbf{k}$ and
physical curvilinear base vectors
$\mathbf{i}_{n}$, $\mathbf{i}_{\tau}$, $\mathbf{i}_{\theta}$
are orthonormalized
$
\iv_n={\mathbf{e}_n}/{\sqrt{g_{nn}}}$,\,
$\iv_\tau={\mathbf{e}_\tau}/{\sqrt{g_{\tau\tau}}}$,\,
$\iv_\ta={\mathbf{e}_\ta}/{\sqrt{g_{\ta\ta}}}$.
The validity of the following relations is
necessary for constucting the contravariant basis
\be
\frac{\p\mathbf{e}_i}{\p x^j}=\frac{\p\mathbf{e}_j}{\p x^i}.
\label{peij}
\ee
The unit vectors of physical coordinate system do not
satisfy such a requirement, in contrast to contravariant basis vectors,
due to the difference in their normalizations.
We note that the requirement~(\ref{peij}) will be satisfied if
the local angle $\w$ between the normal vector to the surface
and the symmetry axis depends only on $x^{\tau}$,
doesn't depend on $x^n$ and $x^\ta$, and
\be
\dfrac{\p\tau}{\p x^\tau}=n\dfrac{d\w}{dx^\tau}+f_0(x^\tau)\, , \quad
\text{i.e.}\qquad
\dfrac{\p^2\tau}{\p x^\tau \p n}=\dfrac{d\w}{dx^\tau}\, .
\label{tauxtauf0}
\ee
Here function $f_0(x^\tau)$ is defined by the geometry
of the problem and, in particular,
by a choice of the origin for the coordinate $n$.
For a spherical surface $x^{\tau}\equiv\w$,
$x^n\equiv n=R\equiv\sqrt{x^2+y^2+z^2}$, $x^\tau\equiv\w=\tau/n$,
$x^\ta\equiv\ta$ and $r=n\sin\w$.

Relations \eqref{tauxtauf0} permit to determine the
contravariant basis near the surface:
${\mathbf{e}_n}={\mathbf{i}_n}$,
$\mathbf{e}_\tau={\mathbf{i}_\tau}{\p \tau}/{\p x^\tau}$,
$\mathbf{e}_\ta=r{\mathbf{i}_\ta}$.
One obtains for the basis the following components of the metric tensor
$g_{nn}=1$, $g_{\tau\tau}=\left({\p \tau/\p x^\tau}\right)^2$,
$g_{\ta\ta}=r^2$, $g_{\tau n}=g_{\tau\ta}=g_{n\tau}=0$,
$g=\det g_{ik}=r^2\left({\p \tau/\p x^\tau}\right)^2$, and
the corresponding Christoffel symbols
\begin{align}
&\Gamma_{n\tau}^{\tau}=\Gamma_{\tau n}^{\tau}=\dfrac{\p\w}{\p x^\tau}
\dfrac{1}{{\p \tau}/{\p x^\tau}}, &
&\Gamma_{n\ta}^{\ta}=\Gamma_{\ta n}^{\ta}=\dfrac{\sin\w}{r} , &
&\Gamma_{\tau\tau}^{n}=-\dfrac{\p\w}{\p x^\tau}
\dfrac{\p \tau}{\p x^\tau} ,
&\Gamma_{\tau\tau}^{\tau}=\dfrac{\p^2\tau}{\p {x^\tau}^2}
\dfrac{1}{{\p \tau}/{\p x^\tau}} , \nonumber\\
&\Gamma_{\tau\ta}^{\ta}=\Gamma_{\ta\tau}^{\ta}=\dfrac{\cos\w}{r}
\dfrac{\p \tau}{\p x^\tau} , &
&\Gamma_{\ta\ta}^{n}=-r\sin\w, & &\Gamma_{\ta\ta}^{\tau}=
-\dfrac{r\cos\w}{{\p \tau}/{\p x^\tau}}
\, .
\label{gammaiik}
\end{align}
The expressions for the metric tensor and Christoffel symbols
allow to obtain explicit formulas for all differential forms,
according to general rules of the differential geometry~\cite{Dubrovin}.
In particular, for arbitrary vector $\mathbf{F}$ one finds
\be
\rot\mathbf{F}=
\dfrac{1}{r}\left(\dfrac{\p \left(r F_\ta\right)}{\p\tau}-
\dfrac{\p F_\tau}{\p \ta}\right)\mathbf{i}_n
+\dfrac{1}{r}\left(\dfrac{\p F_n}{\p \ta}-
\dfrac{\p \left(rF_\ta\right)}{\p n}\right)\mathbf{i}_\tau
+\left(\dfrac{\p F_\tau}{\p n}-\dfrac{\p F_n}{\p\tau}+
\dfrac{d\w}{d\tau}F_\tau\right)\mathbf{i}_\ta\, .
\label{rotntautaphys}
\ee
Therefore,
\bea
\gamma&=&(\rot\vvect)_\ta=\frac{\p v_\tau}{\p n}-
\frac{\p v_n}{\p\tau}+v_\tau \frac{d\varphi}{d\tau},\\
\Delta\vvect&=& -\rot(\gamma\mathbf{i}_{\theta})=
-\mathbf{i}_{n}\left(\frac{\p\gamma}{\p\tau}+\frac{\cos\w}{r}\gamma\right)
+\mathbf{i}_\tau\left(\frac{\p\gamma}{\p n}+\frac{\sin\w}{r}\gamma\right),\\
(\mathbf{v}{\mygrad})\mathbf{v}&=&
\left[v_n\dfrac{\p v_n}{\p n}+
v_\tau\left(\dfrac{\p v_n}{\p \tau}-\dfrac{d\w}{d\tau}
v_\tau\right)\right]\mathbf{i}_n+\left[v_n\dfrac{\p v_\tau}{\p n}
+v_\tau\left(\dfrac{\p v_\tau}{\p\tau}+v_n \dfrac{d\w}{d\tau}\right)\right]
\mathbf{i}_\tau \, .
\eea
Thus, the components of Eq.(\ref{NavSt}) may be rewritten as
\bea
\dfrac{\p p}{\p\tau}=-\rho\left(\dfrac{\p v_\tau}{\p t}+
v_\tau\dfrac{\p v_\tau}{\p \tau}
+v_n\left(\dfrac{\p v_n}{\p \tau}+\gamma\right)\right)
+\eta\left(\dfrac{\p \gamma}{\p n}+\dfrac{\sin\w}{r}
\gamma\right)\, ,
\label{NStau2}
\\
\dfrac{\p p}{\p n}=-\rho\left(\dfrac{\p v_n}{\p t}+
v_n\dfrac{\p v_n}{\p n}
+v_\tau\left(\dfrac{\p v_\tau}{\p n}-\gamma\right)\right)
-\eta\left(\dfrac{\p \gamma}{\p \tau}+\dfrac{\cos\w}{r}
\gamma\right)\, .
\label{nve7}
\eea
In a more general case when $v_\ta\ne 0$, the
terms $\rho v_\ta^2\cos\w/r$, $\rho v_\ta^2\sin\w/r$
should be added to right-hand member of
Eqs.(\ref{NStau2}),(\ref{nve7}) correspondingly.

The components of a viscous stress tensor
$\sigma'_{ik}=\eta(\p v_i/\p x_k+\p v_k/\p x_i)$, which
describe momentum transfer through the boundary, take the form
\be\sigma'_{nn}=2\eta\dfrac{\p v_n}{\p n}\, , \qquad
\sigma'_{n\tau}=\eta\left(\dfrac{\p v_n}{\p\tau}+\dfrac{\p v_\tau}{\p n}
-v_\tau\dfrac{d\w}{d\tau}\right) \, .
\label{sigmannntau}
\ee
The boundary condition at the surface is~\cite{LL6}
\be
\left(P-p_v-\sigma\left(\frac1{R_1}+\frac1{R_2}\right)\right)n_i=
\eta\left(\frac{\p v_i}{\p x_k}+\frac{\p v_k}{\p x_i}\right)n_k-
\frac{\p\sigma}{\p x_i}.
\label{bc1}
\ee
The normal vector is directed here in the outward direction,
towards the atmosphere.
Here $p_v$ a is pressure of the gas and atmosphere, $P$ is a
hydrodynamic pressure on the surface, $R_{1,2}$ are
main local radii of curvature on the surface, and $\sigma$ is
a surface tension. Projections of (\ref{bc1})
to the local tangential and normal directions to the surface are
\bea
\label{Bound}
\frac{d\sigma}{d\tau}&=&
\eta\left(\frac{\p v_n}{\p\tau}+\frac{\p v_\tau}{\p n}-v_\tau\frac{d\w}{d\tau}\right),\\
P-p_v&=&\sigma\left(\frac1{R_1}+\frac1{R_2}\right)+2\eta\frac{\p v_n}{\p n}.
\label{Laplace3}
\eea
Eq.(\ref{Bound}) is the boundary condition for the velocities
on the surface. In particular, one gets the boundary
condition for the vorticity at the surface:
\be
\gamma=\frac1\eta\frac{d\sigma}{d\tau}+
2\left[v_\tau\frac{d\w}{d\tau}-\frac{\p v_n}{\p\tau}\right].
\label{gammaSurf}
\ee
Boundary conditions for stream function at the surface
may be obtained by integrating the expression
${\p\psi}/{\p\tau}=-rv_n(\tau)$, where
$v_n(\tau)$ is the normal component of the velocity to the boundary.
The boundary conditions for the stream function are particulary simple
if the motion of the surface is much slower than
typical fluid velocities of a problem, when one can put $v_n\approx0$.

Eq.(\ref{Laplace3}) represents the boundary condition
that allows to obtain a shape of the surface.
The pressure $P(\tau)$ satisfies Navier-Stokes equations
(\ref{NStau2}), (\ref{nve7}) with corresponding projections of
the gravitational force added to the right-hand part of the
equations. Therefore, the quantity
$p(\tau)=P(\tau)+\rho g z$ satisfies Eqs.(\ref{NStau2}), (\ref{nve7})
without the additional terms.
To further simplify equations and boundary conditions,
one can introduce the quantities
$p_3(\tau)-p_3(0)=p(\tau)-p(0)-\left.2\eta{\p v_n}/{\p n}\right|_0^\tau$,
$p_4(\tau)-p_4(0)=p(\tau)-p(0)+\left.{\rho (v_\tau^2
+v_n^2)}/{2}\right|_0^\tau$, and
$k=R_1^{-1}+R_2^{-1}=d\w/d\tau+\sin\w/r$,
$k_0=\left.(R_1^{-1}+R_2^{-1})\right|_{\tau=0}$, $z_0=\left.z\right|_{\tau=0}$.
Then we get
\bea
\label{p3Shape}
&&p_3(\tau)-p_3(0)=\sigma\left(k-k_0\right)+\rho g(z-z_0),\\
&&\frac{d\w}{d\tau}=k_0+\frac{p_3(\tau)-p_3(0)-\rho g(z-z_0)}{\sigma}
-\frac{\sin\w}{r}.
\label{dwdtau}
\eea
In the particular case when one can disregard the term with the 
pressure, Eq.(\ref{dwdtau}) reduces to the Young-Laplace 
equation in the form obtained in~\cite{RBN}.
The tangential component (\ref{NStau2}) of
the Navier-Stokes equation may be represented as
\be
\frac{dp_4}{d\tau}=
-\rho\left(\frac{\p v_\tau}{\p t}
+v_n\gamma\right)+
\eta\left[\frac{\p\gamma}{\p n}+\frac{\sin\w}{r}\gamma\right].
\label{dp4dtau}
\ee
The equation (\ref{dwdtau}) turns out to be quite convenient
for determining the shape of the surface.
The shape of an axially symmetric surface is
unambiguously described by the function $\w(\tau)$.
Because all the expressions contain either the difference
$p_4(\tau)-p_4(0)$ or the derivative $dp_4(\tau)/d\tau$,
an initial value of $p_4(0)$ is still an arbitrary
constant. It is convenient to take
\be
p_4(0)=2\eta\left.\frac{\p v_n}{\p n}\right|_{\tau=0}+
\left.\frac{\rho (v_\tau^2+v_n^2)}{2}\right|_{\tau=0}.
\label{p4initial}
\ee
Then
\be
p_3(\tau)-p_3(0)=p_4(\tau)-2\eta\frac{\p v_n}{\p n}
-\frac{\rho (v_\tau^2+v_n^2)}{2}.
\label{p3p4a}
\ee
Introducing the vector $\mathbf{y}=(r(\tau),\w(\tau),z(\tau),p_4(\tau))^T$
allows to represent
Eqs. (\ref{dwdtau}),(\ref{dp4dtau}),(\ref{p3p4a}),
\, $dr(\tau)/d\tau=\cos\w$,\,
$dz(\tau)/d\tau=-\sin\w$ in the following form
\be
\frac{d\mathbf{y}}{d\tau}=f(\tau,\mathbf{y}).
\label{ysyst}
\ee
At the initial point one has
$\mathbf{y}(0)=\left(r(0),\w(0),z(0),p_4(0)\right)^T$.
Here $p_4(0)$ is defined in  (\ref{p4initial}).
The Cauchy problem for the system of differential equations
(\ref{ysyst}) with initial conditions derived above can be solved
by standard numerical methods to obtain the surface profile.

{\it Estimations.} It is of interest to find out
a relative role of terms in Eq.(\ref{p3Shape})
under specific physical conditions.
Below we carry out the respective estimatons for an evaporating sessile drop
lying on a substrate in the regime of a pinned contact line.
Evaporation results in an inhomogeneous spatial temperature
distribution in the drop and along the drop surface.
The corresponding Marangoni forces result in
vortex flows of the liquid in the drop.

The motion of the surface is considered to be much slower than
typical fluid velocities.
This property is valid for a wide class of evaporating drops.
Then one can take approximately $v_n\approx 0$. The fluid motion
is considered as a quasistationary vortex flow.
In the following expressions $n_0$ is the characteristic distance between
the surface of the drop and the vortex center,
$r_0$ is the contact line radius, $\sigma'=-\p\sigma/\p T$,
$\Delta T$ is the temperature difference between the substrate
and the apex of the drop, $\theta_c$ is the contact angle.
Therefore, $d\w/d\tau\approx\sin\theta_c/r_0$,
$d\sigma/(\eta d\tau)\approx-\sigma'\Delta T\sin\theta_c/(\eta r_0\theta_c)$.
Using the condition (\ref{Bound}) and taking $n_0$ as a
characteristic distance for a change of $v_\tau$ along the normal
to the surface, one obtains
$v_\tau\approx (\p v_\tau/\p n)n_0=n_0d\sigma/(\eta d\tau)
+n_0v_\tau d\w/d\tau$, i.e.
$v_\tau\left(1-{n_0\sin\theta_c}/{r_0}\right)\approx
-{\sigma'n_0\Delta T\sin\theta_c}/{(\eta r_0\theta_c)}$, hence
\be
|v_\tau| \lesssim \frac{\sigma'\Delta T n_0}{\eta r_0}.
\label{vEstim}
\ee
Therefore
$|v_\tau d\w/d\tau|\approx|v_\tau|\sin\theta_c/r_0\approx
n_0\sigma'\Delta T\sin^2\theta_c/(\eta r_0^2\theta_c)\ll
\sigma'\Delta T/(\eta r_0)$,
i.e. the term $v_\tau d\w/d\tau$ in (\ref{Bound}) and (\ref{gammaSurf})
is much smaller than $d\sigma/(\eta d\tau)$. It follows from
(\ref{gammaSurf}) that
\be
|\gamma|\approx\frac{\sigma'\Delta T}{(\eta r_0)}.
\label{GammaEstim}
\ee
It follows from $|\p^2 v_\tau/\p n^2|\approx |v_\tau|/n_0^2$ and
\be
\frac{\p\gamma}{\p n}=\frac{\p^2 v_\tau}{\p n^2}+
\frac{d\w}{d\tau}\frac{\p\sigma}{\eta\p\tau}
\ee
and (\ref{vEstim}) that
\be
\left|\frac{\p\gamma}{\p n}\right|\approx \frac{\sigma'\Delta T}{\eta r_0 n_0}
\left(\frac{\sin\theta_c}{\theta_c}+\frac{n_0\sin\theta_c}{2r_0}\right)\approx
\frac{\sigma'\Delta T}{\eta r_0 n_0}.
\label{dgdnEstim}
\ee
We substitute (\ref{GammaEstim}) and (\ref{dgdnEstim}) to
(\ref{dp4dtau}) and integrate the obtained expression
over $\tau$. This gives the estimation of relative effects of
pressures and velocities as compared with gravitational forces
in Eq.~(\ref{p3Shape}):
\bea
p_4(\tau)-p_4(0)&\approx&
\frac{\sigma'\Delta T\theta_c}{n_0\sin\theta_c},\qquad\,\,
\left.2\eta\frac{\p v_n}{\p n}\right|_0^\tau \ll p_4(\tau)-p_4(0),
\label{p4Estim}
\\
\frac{|p_4(\tau)-p_4(0)|}{\rho gh}&\approx&
\frac{\sigma'\Delta T}{\rho g n_0 h}\frac{\theta_c}{\sin\theta_c},\qquad
\frac{|\rho v_\tau^2/2|}{\rho gh}\lesssim\frac{1}{2gh}
\left(\frac{n_0}{\eta}\frac{\sigma'\Delta T}{r_0}\right)^2.
\label{p4rhogh}
\eea
For estimating the term $\rho v_\tau^2/2$ we used~(\ref{vEstim}).

The ratio of gravitational force to the term with surface tension
in~(\ref{p3Shape}) is characterised by dimensionless number
$B_0={\rho g h r_0}/{(2\sigma\sin\theta_c)}$, which is analogous
to Bond number. Therefore, (\ref{p4rhogh}) may be represented as
\be
\frac{|p_4(\tau)-p_4(0)|}{|\sigma(k-k_0)|}\approx
\frac{\sigma'\Delta T\theta_c}{2\sigma\sin^2\theta_c}\frac{r_0}{n_0},\qquad
\frac{|\rho v_\tau^2/2|}{|\sigma(k-k_0)|}\approx
\frac{\rho}{4r_0\sigma\sin\theta_c}\left(\frac{n_0\sigma'\Delta T}{\eta}\right)^2.
\ee

{\it Conclusion.} Based on a geometry of the fluid surface, we
have derived Eqs.(\ref{NStau2}),(\ref{nve7})
and the boundary condition (\ref{p3Shape}), which allow to obtain
numerically a surface profile dynamics and to take into
account the influence of fluid dynamics and gravitational
forces on the shape of the fluid-gas interface.

The equations and boundary conditions derived in this paper
were used in~\cite{Drop} to find numerically the profile
of the evaporating sessile drop surface.
According to Eq.~(\ref{p4rhogh}), the effects of the pressure 
become more important with the increase of the temperature drop 
in the liquid and with the temperature derivative of the surface tension.
Analytical estimations~(\ref{p4rhogh}) applied
to the conditions of~\cite{Drop} show that the relative
contribution of pressures and velocities as compared with
gravitational forces in the Laplace formula~(\ref{p3Shape}), 
is not too large. 
Numerical results confirm this qualitative conclusion
and give approximately one tenth for the value of this quantity
under the conditions of~\cite{Drop}.

The author is grateful to V.V.~Lebedev and L.N.~Shchur
for useful discussions and remarks.


\begin{thebibliography}{99}

\bibitem{Deegan97} R.~D.~Deegan et al., Nature {\bf 389}, 827 (1997).

\bibitem{Deegan} R.~D.~Deegan et al., Phys. Rev. E {\bf 62}, 756 (2000).

\bibitem{HuLarsonEvap} H.~Hu, R.~G.~Larson, J. Phys. Chem. B
{\bf 106}, 1334 (2002).

\bibitem{Popov1} Y.~O.~Popov and T.~A.~Witten, Phys. Rev. E
{\bf 68}, 036306 (2003).

\bibitem{Ristenpart} W.~D.~Ristenpart, P.~G.~Kim, C.~Domingues,
J.~Wan, H.~A.~Stone, Phys. Rev. Lett. {\bf 99}, 234502 (2007).

\bibitem{HuLarsonMarangoni} H.~Hu, R.~G.~Larson, Langmuir {\bf 21}, 3972 (2005).

\bibitem{Girard} F.~Girard, M.~Antoni, S.~Faure, A.~Steinchen,
Langmuir {\bf 22}, 11085 (2006).

\bibitem{Dhaval} 
H.~K.~Dhavaleswarapu, P.~Chamarthy, S.~V.~Garimella,
J.~Y.~Murthy, Phys. Fluids {\bf 19}, 082103 (2007).

\bibitem{Govor} L.~V.~Govor, G.~Reiter, J.~Parisi, G.~H.~Bauer,
Phys. Rev. E {\bf 69}, 061609 (2004).

\bibitem{HuLarsonCoffee} H.~Hu, R.~G.~Larson, J. Phys. Chem. B,
{\bf 110}, 7090 (2006).

\bibitem{Popov2} R.~Zheng, Y.~O.~Popov, T.~A.~Witten,
Phys.Rev. E {\bf 72}, 046303 (2005).

\bibitem{Jing} J.~P.~Jing et. al., Proc. Natl. Acad. Sci. U.S.A. {\bf 95}, 8046 (1998).

\bibitem{HuLarsonDNA} M.~Chopra, L.~Li et. al., J. of Rheology {\bf 47}, 1111 (2003).

\bibitem{Hsieh} C.~Hsieh, L.~Li, R.~G.~Larson, 
J. Non-Newtonian Fluid Mech {\bf 113}, 147 (2003).

\bibitem{Park} J.~Park, J.~Moon, Langmuir {\bf 22}, 3506 (2006).

\bibitem{Jong} J.~Jong et. al., Appl. Phys. Lett. {\bf 91}, 204102 (2007).

\bibitem{Lim} J.~Lim et. al., Adv. Funct. Mater. {\bf 18}, 229 (2008).

\bibitem{Lin1} X.~M.~Lin, H.~M.~Jaeger, C.~M.~Sorensen, K.~J.~Klabunde,
J. Phys. Chem., B {\bf 105}, 3353 (2001).

\bibitem{Lin2} S.~Narayanan, J.~Wang, X.~M.~Lin,
Phys. Rev. Lett. {\bf 93}, 135503 (2004).

\bibitem{Bigioni3} T.~P.~Bigioni, X.~M.~Lin, T.~T.~Nguyen,
E.~I.~Corwin, T.~A.~Witten, H.~M.~Jaeger,
Nature Materials {\bf 5}, 265 (2006).

\bibitem{LL6} L.~D.~Landau and E.~M.~Lifshitz, {\it
Course of Theoretical Physics VI: Fluid Mechanics}
(Pergamon Press, Oxford, 1982).

\bibitem{Dubrovin}
B.~A.~Dubrovin, A.~T.~Fomenko and S.~P.~Novikov, {\it Modern
Geometry - Methods and Applications} (Springer-Verlag, New
York/Berlin, 1984).

\bibitem{RBN} Y.~Rotenberg, L.~Boruvka, A.~W.~Neumann,
J. Coll. Int. Sci. {\bf 93}, 169 (1983).

\bibitem{Drop}
L.~Yu.~Barash, L.~N.~Shchur, V.~M.~Vinokur, T.~P.~Bigioni, H.~M.~Jaeger,
in preparation.

\end{thebibliography}
\end{document}